\documentclass[
  journal=largetwo,
  manuscript=article-type,
  year=2020,
  volume=37,
]{cup-journal}

\usepackage{amsmath}
\usepackage[nopatch]{microtype}
\usepackage{booktabs}
\usepackage{graphics}
\usepackage{listings}
\usepackage{pgfplots}
\usepackage{algorithm2e}
\usepackage{subcaption}
\usepackage{multicol}
\usepackage{xcolor}
\usepackage{blindtext}
\usepackage{tablefootnote}

\title{BLINK: an End-To-End GPU High Time Resolution Imaging Pipeline for Fast Radio Burst Searches with the Murchison Widefield Array}

\author{Cristian Di Pietrantonio}
\affiliation{Pawsey Supercomputing Research Centre, Kensington, 6151, Western Australia, Australia}
\alsoaffiliation{International Centre for Radio Astronomy Research, Curtin University, Bentley, 6102, Western Australia, Australia}
\email[Cristian Di Pietrantonio]{cristian.dipietrantonio [at] csiro.au}

\author{Marcin Sokolowski}
\affiliation{International Centre for Radio Astronomy Research, Curtin University, Bentley, 6102, Western Australia, Australia}

\author{Christopher Harris}
\affiliation{Pawsey Supercomputing Research Centre, Kensington, 6151, Western Australia, Australia}

\author{Daniel Price}
\affiliation{International Centre for Radio Astronomy Research, Curtin University, Bentley, 6102, Western Australia, Australia}
\alsoaffiliation{Square Kilometre Array Observatory (SKAO), Kensington, 6151, Western Australia, Australia}

\author{Randall Wayth}
\affiliation{International Centre for Radio Astronomy Research, Curtin University, Bentley, 6102, Western Australia, Australia}
\alsoaffiliation{Square Kilometre Array Observatory (SKAO), Kensington, 6151, Western Australia, Australia}


\begin{document}

\begin{abstract}

Petabytes of archival high time resolution observations have been captured with the Murchison Widefield Array. The search for Fast Radio Bursts within these using established software has been limited by its inability to scale on supercomputing infrastructure, necessary to meet the associated computational and memory requirements. Hence, past searches used a coarse integration time, in the scale of seconds, or analysed an insufficient number of hours of observations. This paper introduces BLINK, a novel radio interferometry imaging software for low-frequency FRB searches to be run on modern supercomputers. It is implemented as a suite of software libraries executing all computations on GPU, supporting both AMD and NVIDIA hardware vendors. These libraries are designed to interface with each other and to define the BLINK imaging pipeline as a single executable program. Expensive I/O operations between imaging stages are not necessary because the stages now share the same memory space and data representation. BLINK is the first imaging pipeline implementation able to fully run on GPUs as a single process, further supporting AMD hardware and enabling Australian researchers to take advantage of Pawsey's Setonix supercomputer. In the millisecond-scale time resolution imaging test case illustrated in this paper, representative of what is required for FRB searches, the BLINK imaging pipeline achieves a 3687x speedup compared to a traditional MWA imaging pipeline employing WSClean.

\end{abstract}

\section{Introduction}

 Fast Radio Bursts (FRBs) are astronomical radio transients occurring in the millisecond timescale and releasing a large amount of energy, of the order of $10^{39}$\, erg, across a wide range of radio frequencies. Their origin is yet to be confirmed. FRBs travel cosmological distances to reach Earth, implied by their large dispersion measure (DM) which can be in the thousands pc\,cm$^{-3}$, and they come from directions distributed uniformly at random across the entire sky \autocite{frb}.  The vast majority of detections occurred between frequencies $350$\,MHz and $1.8$\,GHz mainly using the Canadian Hydrogen Intensity Mapping Experiment (CHIME) and Australian Square Kilometre Array Pathfinder (ASKAP) telescopes. Low frequency FRBs are much rarer \autocite{Pleunis2021, Gopinath2023, Tyulbashev2024}. 
 
 The Murchison Widefield Array (MWA) is a low-frequency radio telescope located in Western Australia \autocite{Wayth2018,mwa2013}. It observes at frequencies in the 70-300\,MHz range and it is able to instantly capture a frequency bandwidth of 30.72 MHz, divided in 24 coarse channels. Each coarse channel may be fine channelised on site by FPGAs. Phase I of the instrument presented 128 tiles of 16 dipole antennas. Phase II expanded the number of tiles to 144. The MWA is equipped with a Voltage Capture System (VCS; \cite{Tremblay2015}) to save voltage data recorded from its tiles at 100\,$\mu$s time and 10\,kHz frequency resolutions. It grants researchers freedom to employ custom computational techniques to detect signals with very short duration at the cost of dealing with a much larger volume of data. Pulsars \autocite{Bhat2016, Meyers2018, Sett2023, Bhat2023, Bhat2023a, Grover2024} have been found using the MWA VCS given the high time and frequency resolutions required to detect them.
 
 The MWA, with its high angular resolution and large Field-of-View (FoV), could detect the first FRBs at low frequencies in the southern sky. Radio astronomers have been searching for low-frequency FRBs with the MWA adopting beamforming \autocite{Tian2022} and imaging \autocite{Tingay2015,Sokolowski2018,Kemp2024} techniques without finding any. 
 
 Detection of FRBs at low frequencies is computationally challenging. It requires processing data at high resolution in time and frequency to avoid smearing the signal, and in longer time intervals due to quadratic signal dispersion across frequency channels. Coupled with the large number of receivers, wide bandwidth, and extended sky coverage that characterise modern radio telescopes such as the MWA, these requirements translate into high data rates difficult to manage. 

In this work we present the Breakthrough Low-latency Imaging with Next-generation Kernels (BLINK) pipeline, a software suite to execute radio astronomy imaging at high time and frequency resolution, in the order of 20\,ms and 40\,kHz respectively, starting from high time resolution voltages recorded with the MWA. The ultimate science goal is to look for, detect, and localise FRBs within 3\,PB of archival MWA VCS observations using an imaging-based search pipeline. Hence, the implementation prioritises processing efficiency by avoiding unnecessary computational steps and speeding up the execution of fundamental ones through end-to-end accelerated computing on Graphics Processing Unit (GPU) hardware. The implied computational power required to perform high time resolution imaging on the petabyte-sized MWA archive underpins the choice of Setonix \autocite{setonix}, currently Australia's most powerful public research supercomputer, as the target system for execution.

This paper is structured as follows. In Section \ref{sec:related_work} we explore other imaging pipelines that have been used to look for fast transients, especially in MWA data, and in Section \ref{sec:motivation} we motivate the need to develop a completely new one. Section \ref{sec:imaging_basics} provides the theoretical basics of the imaging process, and why it is preferred to beamforming. Section \ref{sec:smart_pipeline} describes the traditional MWA imaging pipeline and its shortcomings when high frequency and time resolutions are demanded. Section \ref{sec:blink_pipeline} introduces the BLINK imaging pipeline developed as part of this work. Section \ref{sec:testing} demonstrates that images produced by BLINK are correct and reasonably close to the ones generated using the standard tool WSClean. It also demonstrates the detectability of transients within them at millisecond-scale time resolution. Section \ref{sec:benchmarking} shows the computational advantage in using BLINK over a WSClean-based solution to look for FRBs. Finally, we present our conclusions in Section \ref{sec:discussion} and plans for future work in Section \ref{sec:future_work}.

\section{Related work} \label{sec:related_work}


The vast majority of FRBs have been detected using beamforming techniques and dedicated compute infrastructure \autocite{Amiri2021}.  The CHIME/FRB collaboration established an FRB detection system that upgrades the GPU correlator of the CHIME telescope to produce 16,384 fine channels at 1\,ms time resolution. The correlation software runs on 256 dedicated processing nodes, each with two commercial dual-chip AMD GPUs. The output is fed to a beamforming pipeline creating 1024 beams across the sky, then passed to further stages for detection and classification of FRBs \autocite{Amiri2018}.

However, imaging is emerging as an alternative method to recover the sky intensity over time and frequency. The ASKAP's CRAFT project presented the CRAFT COherent upgrade (CRACO), an FRB search system based on millisecond-scale imaging implemented on 22 FPGA boards. By forming 256x256 images covering 1.1 square degrees of sky, their goal is to obtain precise localisation of detected FRB events \autocite{Wang2025}.

Observations taken with the MWA have been searched several times for FRBs \autocite{Tingay2015, Rowlinson2016, Sokolowski2018, Kemp2024} using imaging-based techniques. For instance, the MWA was used to shadow the ASKAP telescope by recording and imaging visibilities at 0.5\,s time and 1.28\,MHz frequency resolutions. While ASKAP detected FRB pulses, no low-frequency counterparts were found \autocite{Sokolowski2018}. If bright enough FRBs were indeed recorded by the MWA, likely reasons for null results would be coarse integration times, in the scale of seconds, insufficient hours of observations analysed based on the expected FRB rate \autocite{Sokolowski2024}, or both.

All MWA imaging pipelines are a sequence of established radio astronomy packages whose execution is orchestrated using BASH or Nexflow. COTTER \autocite{cotter} handles visibilities correction and data format conversion from FITS to MeasurementSet; AOFlagger applies RFI mitigation; WSClean \autocite{Offringa2014}, Miriad or CASA are adopted for gridding and imaging. Each step in the pipeline reads in from the filesystem the previous step's output and produces files to be given as input to the next one. An in-depth description of a representative MWA imaging pipeline is given in Section \ref{sec:smart_pipeline}. 

Except for \citet{Kemp2024}, none of the highlighted MWA publications mentions the computational setup and costs of running such searches. In the latest search for FRBs within more than one hundred hours of MWA observations, \citet{Kemp2024} underline the importance of parallel processing and scalability to reduce compute time by taking advantage of supercomputing infrastructure. In their approach multiple 2-minute observations are distributed across and independently imaged at 2 second resolution by 64-core CPU nodes of a commercial cluster. The Input / Output (I/O) bottleneck afflicting established MWA processing software is addressed by mounting RAM as a filesystem where intermediate files are written to and read from. By doing so, they managed to process 400, 2-minute observations in 25 minutes.

 The first GPU implementation of voltage correlation was presented by \citet{Harris2008}, a work which demonstrated its feasibility and computational advantage over a CPU implementation. The research fed into a first GPU correlation system developed for the MWA prototype \autocite{Randall2009GPU} as well as into the highly optimised xGPU correlation library \autocite{Clark2011}. The latter has been subsequently adopted for the MWA online correlation system \autocite{correlator, Morrison2023}. \citet{Romein2021}  presented a GPU correlator that leverages the NVIDIA Tensor Cores to achieve up to 10x speed up compared to xGPU. The idea of an imaging pipeline for the MWA taking advantage of the computational power of GPUs is not new \autocite{Ord2009}, but has not effectively taken over the established workflows mostly based on CPU software. GPU acceleration is only routinely used through the Image Domain Gridder (IDG) software \autocite{2019ascl.soft11011V}, part of the popular WSClean imaging application. 

RICK is a GPU software with a similar philosophy to BLINK but a different objective \autocite{DeRubeis2025}. It is a library of imaging routines developed to run on HPC environments with MPI support and NVIDIA GPU acceleration. They focus on the processing of LOFAR massive visibilities datasets to produce large images, of the order of 500\,GB and 65K$\times$65K resolution, respectively.

The BLINK GPU imager is a software designed to execute visibilities corrections, gridding, and FFT, making up the later part of the imaging process, entirely on GPU, supporting both NVIDIA and AMD vendors \autocite{GPUImager}. It specifically targets the scientific goal of finding FRBs and hence focusing on meeting the demanding computational requirements imposed by high time and frequency resolutions, and a large FoV. The BLINK GPU imager is a component of the BLINK imaging pipeline presented in this work, and the BLINK FRB search pipeline, to be implemented in the future.

\section{Motivation} \label{sec:motivation}







\begin{table}
	\begin{tabular}{|l|l|}
		\toprule
		\textbf{Property} & \textbf{Value} \\
		\bottomrule
		Number of tiles & 128 \\\hline
		Central Frequency & 150\,MHz \\\hline
		Bandwidth & 30.72\,MHz \\\hline
		Number of fine channels & 768 \\\hline 
		Fine channel resolution & 40\,kHz \\\hline 
		Integration time & 50\,ms \\\hline
		Image size & 512x512\,px \\\hline
		Dispersion Measure & 600\,pc\,cm$^{-3}$ \\\hline
		Dispersive delay & 40\,s \\\hline
		Number of Images per time frame & $\approx$ 600k \\\hline
		Data volume per time frame & $\approx$ 584 GiB \\\hline
	\end{tabular}
	
	\caption{\textbf{Representative requirements and constraints for an FRB search with the MWA.} The high time and frequency resolution requirements, the large instantaneous bandwidth of the MWA, and the dispersive delay causing the signal to cross the low frequency band in tens of seconds demand the FRB search software to handle half a million of images per each time frame needed to compute dynamic spectra.}
	\label{tab:frb_requirements}
\end{table}

 A summary of representative science requirements imposed on FRB searches with the MWA is presented in Table \ref{tab:frb_requirements}, informed by the telescope specifications and other FRB searches in similar settings \autocite{Pleunis2021,Gopinath2023}. FRBs are subject to large DM and accentuated dispersion at low-frequencies, meaning that tens of seconds must be considered when constructing dynamic spectra. The dispersive delay is three orders of magnitude larger than the integration time. This results in more than half a million of images, which occupy at least 0.5\,TiB of memory, being manipulated roughly at any given time during the processing.

Then, processing tens of hours worth of observations in a timely and efficient way requires more computational power, measured in FLoating-point OPerations per Second (FLOPS), than a single workstation or a small cluster can provide. The new Setonix supercomputer \autocite{setonix}, hosted at the Pawsey Supercomputing Research Centre, provides a partition of GPU nodes comprising 768 AMD MI250X GPUs that reaches approximately $35$\,PFLOPS, representing the majority of compute provided by the entire system. In fact, nearly all the newly-procured supercomputers derive most of their compute power from GPU accelerators \autocite{top500gpu} due to their computational speed and energy efficiency.

However, the vast majority of radio astronomy software developed by researchers is designed to run in isolation on CPU-only workstations or small clusters at most.  Only a small subset of established and widespread software leverage GPUs, and they target NVIDIA ones, making the majority of Setonix compute power inaccessible to these applications. Furthermore, even when GPUs are used, the extensive reliance on the filesystem to store intermediate data products considerably slows down the processing. In recent years, radio astronomers and developers have started addressing the aforementioned issues but have not converged to a solution yet \autocite{Edgar2010, rts, Law2018}.

The BLINK imaging pipeline is a new radio astronomy software written with the primary goal of running large scale FRB searches in the MWA archival data. All computations are executed on GPU and intermediate data products reside in GPU memory, removing unnecessary I/O operations. The software is written to support both NVIDIA and AMD GPUs. While the BLINK imaging pipeline has been successfully tested on an NVIDIA cluster, it is mainly developed on the Setonix GPU partition, its target execution environment. By using Setonix, a large number of observations can be processed in parallel over multiple GPUs and compute nodes, drastically increasing the FLOPS available to the computation.

\section{The Imaging Process}\label{sec:imaging_basics}


Interferometers feature a large number $N_a$ of independent antennas spread over a significant surface area. Each antenna records time series $V_a(t)$ of voltage data describing radiation coming from a range of frequencies and from all directions within the antenna beam, corresponding to the FoV. There are different ways of combining these signals into a robust source of information. In coherent beamforming, signals are coherently added together to form a voltage beam $b(t)$:

\begin{equation}\label{eq:beamforming}
	b(t) = \sum_{a = 1}^{N_a} w_a V_a(t),
\end{equation}

where $w_a$ is a per-antenna correction term used to phase signals to point towards a specific direction in the sky. The power beam $B$ is obtained by squaring and averaging in time:
\begin{equation}
	B = \left\langle b(t) b^*(t) \right\rangle.
 \end{equation}

The computational time complexity of beamforming the entire FoV is $\Theta(N_a N_x^2)$, where $N_x = B_\textit{max} / D$, $B_\textit{max}$ is the maximum distance between antennas, and $D$ is the diameter of an antenna aperture. 

As the angular resolution and FoV of modern radio telescopes increase, beamforming becomes computationally prohibitive. Imaging is a viable alternative to analyse data from low-frequency radio telescopes such as the MWA and Low Frequency Square Kilometre Array \autocite{skadesc}, although still very computationally expensive.

In the interferometric imaging process \autocite{Condon2016}, The first step is to construct the cross power spectrum by multiplying fine-channelised voltages from each pair of antennas $(a_i, a_j)$, and averaging the result in time \autocite{correlator}:

\begin{equation}\label{eq:correlation}
	S^\nu_{a_i, a_j} = \left\langle \int_{-\infty}^{\infty} V_{a_i} (t) e^{2\pi i\nu t} dt \cdot  \int_{-\infty}^{\infty} V^*_{a_j} (t) e^{2\pi i\nu t} dt \right\rangle.
\end{equation}

The fine channelisation has a cost of $\Theta(N_aF\log_2F)$ and the multiplication is done in $\Theta(N_a^2)$ operations.
 
The cross power spectrum $S^\nu_{a_i, a_j}$ measures a finite sample of the spatial frequency space  $S^\nu(u, v)$ of the radio signal, also called visibility function. Let $(u, v)$ be the coordinates in such space, and $(l, m)$ the direction cosines of sky pointings. Then, visibilities are related to the sky brightness distribution by the following approximation:

\begin{equation}\label{eq:visibility}
	S^\nu(u, v) = \int\int \frac{A(l, m) I^\nu(l, m)}{\sqrt{1 - l^2 -m^2}}e^{-i2\pi(ul + vm)}\, dl \, dm,
\end{equation}

where $A(l, m)$ is the primary beam of the telescope, and $I^\nu(l, m)$ is the sky specific intensity one wishes to measure. Equation \ref{eq:visibility} defines the visibility function as Fourier transform of the sky intensity. 

Samples in $S^\nu_{a_i, a_j}$ are placed on a regular grid $S^\nu(u_x, v_y)$ whose side turns out to be of length $N_x$, in a process called gridding. Then, the 2D inverse Fast Fourier Transform (2D FFT) algorithm is used to recover an approximation of $I^\nu(l, m)$, the dirty image  $\hat{I}^\nu(l, m)$:

\begin{equation}
  \hat{I}^\nu(l, m) =	2D FFT(S^\nu(u_x, v_y)).
\end{equation}

 It is an approximation because the finite number of antenna pairs does not provide full coverage of the $(u, v)$ plane. The computational time complexity of gridding is $\Theta(N_a^2)$, and the 2D FFT one is equal to $\Theta(N^2_x\log(N^2_x))$.

Most science applications of imaging require a further step named deconvolution, where artifacts due to the discrete sampling of the $(u, v)$ plane are removed  from the dirty image. CLEAN \autocite{Hogbom1974} is an iterative algorithm that implements deconvolution and it is computationally expensive to run. However, FRBs are so bright that such distortion effects do not practically impair the detection process and the CLEAN step is not necessary. For the same reason, we do not correct distortions caused by the wide FoV.

\begin{figure}
	\centering
	\includegraphics[width=\linewidth,]{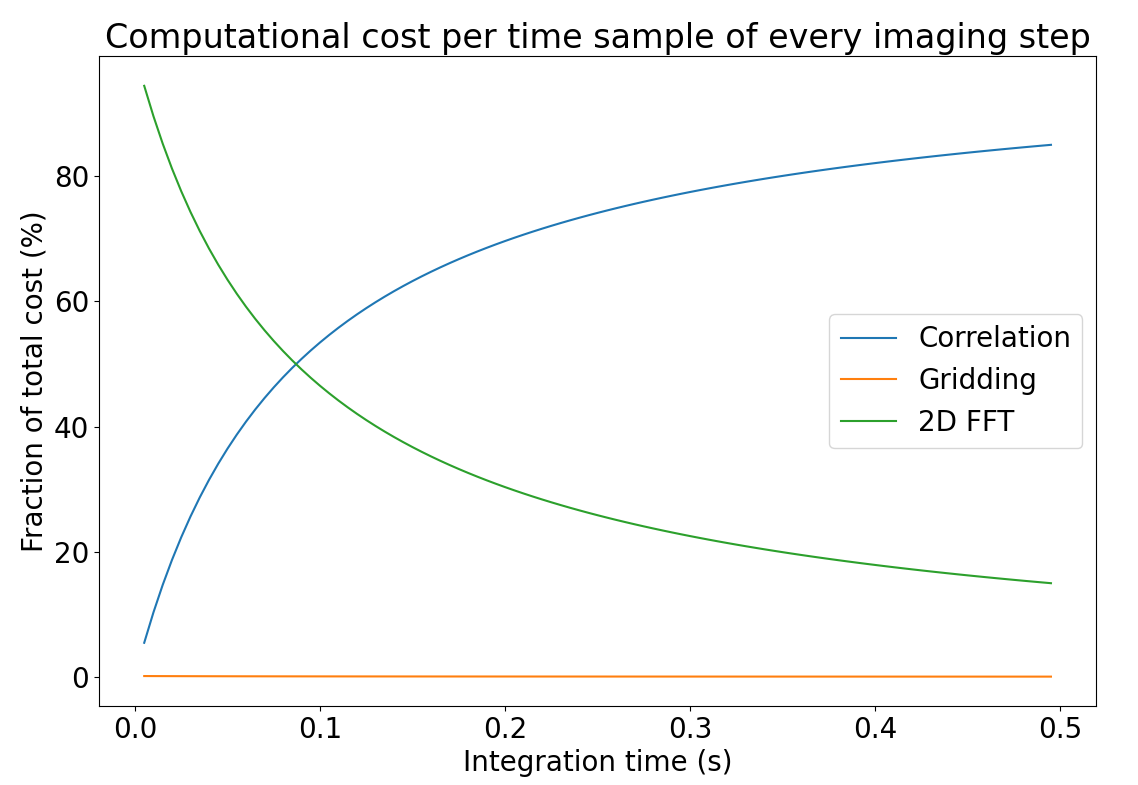}
	\caption{\textbf{Relative computational cost per time sample of the main steps of the imaging process.} The number of fundamental numerical operations of the correlation, gridding and 2D FFT algorithms expressed as a percentage of their sum is a proxy of their relative computational cost. Curves shown in the plot arise when the imaging process is instantiated on MWA Phase I VCS data at observational frequency of 150MHz. The integration time is the critical parameter that determines whether 2D FFT or correlation is the dominant part of the computation.}
\label{fig:imaging_costs}
\end{figure}

\begin{figure}
	\centering
	\includegraphics[width=\linewidth,]{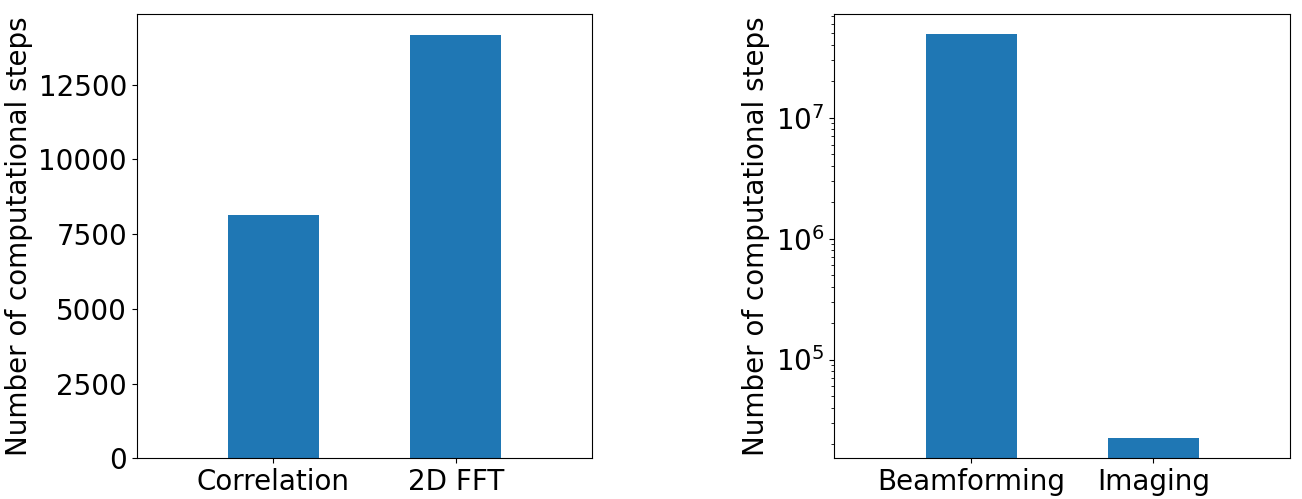}
	\caption{\textbf{Significant computational costs in the context of an FRB search.}The computational cost per time sample of correlation and 2D FFT in the imaging process for the MWA Phase I VCS data, at an integration time of 50\,ms and observational frequency of 150\,MHz (left panel). In the same setting, beamforming requires three orders of magnitude more computation than imaging (right panel).}
	\label{fig:bf_vs_img}
\end{figure}

Correlation and inverse 2D FFT are the most computationally expensive steps of the dirty imaging process. Assuming the image side length $N_x$ is greater than the number of antennas $N_a$, the execution of the correlation algorithm is likely to be faster than the 2D FFT one. However, the former is performed at every time sample whereas the latter only at the end of an integration interval containing $N_s$ time samples. The cost of the 2D FFT per time sample is $\Theta(N^2_x\log(N^2_x)N_s^{-1})$. Figure \ref{fig:imaging_costs} illustrates the contribution of correlation, gridding and 2D FFT to the total execution time of the imaging process as a function of the integration time. Figure \ref{fig:bf_vs_img} shows that correlation and 2D FFT are almost equally important for a representative integration time required to study FRBs.

The reason for choosing imaging over beamforming is the computational cost. If correlation can be, within a constant factor, considered as representative of the cost of the imaging process, then the ratio between beamforming and imaging costs is $\Theta(N^2_x / N_a)$. Because the number of pixels is at least three orders of magnitude larger and increases much faster than the number of antennas, the growth of the ratio highlights the expensive cost of beamforming. The advantage of imaging can also be found with a more precise analysis on the number of computational steps that considers the time complexity of correlation and 2D FFT, presented in the right panel of Figure \ref{fig:bf_vs_img}.

\section{Traditional MWA Imaging Pipeline}\label{sec:smart_pipeline}

The imaging pipeline described by \citet{Sett2023} is a representative example of how MWA observations are traditionally imaged. Because authors in the referenced paper use it to analyse the SMART survey \autocite{Bhat2023, Bhat2023a}, we will refer to it as the SMART pipeline. It starts by correlating voltages captured by the MWA VCS using the offline correlator program. The resulting visibilities are saved in FITS files, one for each coarse channel. Internally, offline correlator uses the GPU-accelerated xGPU library. While still retaining high performance on modern NVIDIA GPUs, xGPU was developed more than ten years ago, and it has not been maintained for six. The code contains extensive low-level optimisation logic that makes any porting to AMD GPUs unfeasible. There is another practical limitation in using xGPU: parameters such as integration time and number of antennas must be chosen at compile time. Multiple xGPU deployments are required for observations taken with different configurations of the MWA, introducing a significant administrative overhead.

Before they can be processed into science-ready data products, visibility data are corrected for instrumental and environmental effects by application of calibration, and radio frequency interference (RFI) by flagging. This is partially accomplished by multiplying each visibility point by a calibration solution that is determined using a reference object in the sky observed before or after the main observation. Other corrections involve applying a phase shift to the signal to digitally steer the telescope. These operations collectively form the preprocessing step and they are implemented in the purpose-built software COTTER. It interleaves I/O and computation, and speeds up the whole execution using CPU multi-threading. While these corrections are applied relatively quickly, most of the time is spent in I/O. The corrected visibilities are stored on the filesystem using the CASA MeasurementSet format.

WSClean is the standard software used to image visibilities. It implements a wide variety of functionalities and it is named after its implementation of the w-stacking technique to image wide FoVs, and its CLEAN implementation. For the purpose of an image-based FRB search, only the gridding and the 2D FFT are needed to produce dirty images. Within gridding, visibilities are convolved with a gridding kernel to reduce aliasing effects of out-of-FoV sources. The architecture of WSClean relies extensively on the CASA library, and in particular the \texttt{MeasurementSet} class to read input visibilities, to hold data through the various computational steps, and to write the resulting images back to disk. The computation takes place mainly on CPU. The gridding step is sped up with GPU acceleration through the Image Domain Gridder (IDG) plug-in module.

Automation in this setting is achieved by scripting invoking the offline correlator, COTTER, and WSClean programs for each second of the observation. The output are multiple MeasurementSet directories and FITS files representing intermediate and final data products.

\section{BLINK Pipeline implementation} \label{sec:blink_pipeline}

This paper introduces the Breakthrough Low-latency Imaging with Next-generation Kernels (BLINK) pipeline to efficiently image voltage capture data in high time and frequency resolution. It will make up the first half of a larger pipeline designed to look for low-frequency FRBs within MWA observations. The code repositories for the BLINK project are publicly available on GitHub at the following link: \url{https://github.com/PaCER-BLINK-Project}.

\subsection{Overview}

\begin{figure*}
	\includegraphics[width=0.8\linewidth]{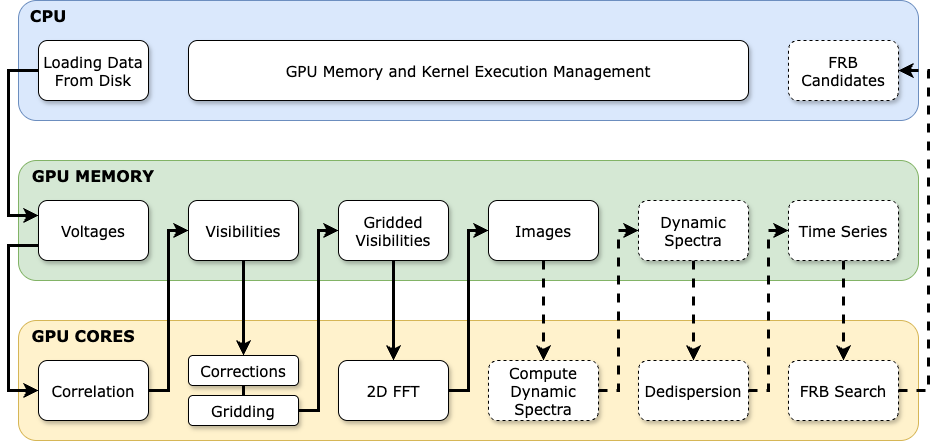}
	\caption{\textbf{The BLINK FRB search pipeline}. The distinguishing features of the BLINK pipeline are the utilisation of GPU cores to execute every compute stage and a homogenous data representation throughout. In addition to parallelisation of computation, the proposed approach eliminates the need of using the filesystem as a temporary staging storage. Intermediate data products to feed successive stages already reside in GPU memory, removing expensive CPU-GPU memory transfers. The CPU is left the task of reading the voltage data from disk, planning and submitting GPU computations, managing memory, and presenting the final output to the user. Dashed lines characterise steps yet to be implemented.}
	\label{fig:frb_pipeline}
\end{figure*}

BLINK is a new software suite written in C++ for which computation exclusively occurs on GPU hardware, where it is parallelised across hundreds of compute units. GPU acceleration is preferred over multi-core CPUs as the former delivers a larger quantity of FLOPS than the latter, and does so more efficiently in terms of energy consumption. Visibilities and images are encoded using numerical values in 32 bit floating point format and they are manipulated with the same precision.  A schematic representation of the BLINK pipeline is shown in Figure \ref{fig:frb_pipeline}. No compute operation is performed on the CPU, which is now dedicated to initial I/O operations and GPU management. Data at each stage of the computation reside in GPU memory, ready to be processed by the next stage. An observation can be partitioned into independent time intervals and processed by multiple GPUs. Given the high data volumes involved, as shown in Figure \ref{fig:data_rates}, the distribution of work also ensures each GPU is assigned a manageable amount of data to process. The software currently implements the first half of the illustrated process, up to the generation of images. Future work, illustrated with dashed lines, will develop the FRB search highlighted in the second half. Ultimately, only the final FRB candidates identified by the software will be transferred to CPU memory and saved to disk for inspection and archiving.

Stages of the imaging process are implemented as independent but interoperable C++ libraries whose routines are then used to develop an imaging pipeline as an end-user executable computer program. The modularity of the design encourages code reusability and generalisation to other science cases and interferometers.

\subsection{Computational steps}




\subsubsection{Data ingestion}  

Voltages collected by the MWA VCS are archived at Pawsey either on the Banksia tape storage or Acacia object store system. A VCS observation is then accessible on-demand and can be transferred on Setonix's scratch Lustre filesystem for processing on the supercomputer. 

An MWA VCS observation can come in different file formats and state of readiness depending on the telescope configuration at the time voltage data were recorded. Early preprocessing stages such as fine channelisation and data reordering may be required. Because the focus is on the imaging process, this work assumes MWA Phase I voltage data format, which is already fine-channelised and stored in binary files, one for each coarse channel and for each second of observation. 


A second of observation consists of 24 binary \texttt{.dat} files, one per coarse channel. Each file contains complex voltage data divided into 128, 10\,kHz fine channels. An instance of the BLINK pipeline processes a second of data at a time by reading in parallel all 24 \texttt{.dat} files in buffers of Managed Memory allocated on the host but accessible to GPUs. Rather than copying data with one single transfer, or explicitly using streams to overlap computation and data movement, such approach allows the GPU driver to asynchronously and concurrently transfer data as they are accessed during the GPU kernel execution.




\subsubsection{Correlation}

During correlation voltage data from every pair of antennas are combined to produce visibilities by implementing Equation \ref{eq:correlation}. In practice, the MWA VCS data have already been fine channelised using a FPGA-based system. Hence, only the Conjugate Multiply Accumulate (CMAC) operation was developed. 

Two levels of parallelisation are in place. At the outer level pairs of voltage time series, one pair for each baseline, frequency channel and integration interval, are assigned to groups of 32 or 64 threads executed in lockstep called warps. Warps execute in parallel over all Compute Units (CU) of the GPU. Within each warp, and for each polarisation, threads compute in parallel the CMAC operation on independent subsets of the time series. These partial results are summed together through a parallel reduction operation spanning all threads in the warp, and the result is divided by the integration time and number of averaged frequency channels. 

Visibilities are stored in a lower triangular matrix form for each integration interval and output frequency channel.








\subsubsection{Visibilities corrections}

Visibilities generated by the correlation routine undergo a series of preprocessing steps before they are fed into the imaging stage. All of the following corrections except reordering modify the visibility matrix in place, avoiding unnecessary memory consumption and maximising cache usage. Visibilities are one-to-one mapped to GPU threads that process them independently in parallel, the simplest form of GPU parallelism.

First, data are reordered such that the index of each row and column in the correlation matrix corresponds to the ordinal number of the corresponding antenna. This is not already the case due to implementation details of the fine channelisation stage.  Next, precomputed calibration solutions are applied to visibilities, followed by geometric and cable length corrections. Cable length correction accounts for the different lengths of on-site cables transmitting voltages from antenna stations to receivers. Geometric corrections set the phase centre to a desired pointing direction. This last step is currently implemented in the imaging library where the required $(u, v, w)$ coordinates are computed.

\subsubsection{Imaging}

The BLINK pipeline adopts the GPU imager presented in \autocite{GPUImager} to recover the sky brightness by executing the gridding and inverse FFT operations. The GPU imager, also developed within the BLINK project, takes as input the visibilities already residing on GPU memory, configuration parameters such as image size and metadata information it uses to compute $(u, v, w)$ coordinates of every baseline, among other necessary information. Compared to the original version, the imager has been further improved such that gridding tasks run in parallel on the same GPU to process multiple fine channels and integration intervals. Similarly, all of the grids are then fed at once to the inverse 2D FFT GPU implementation  for batch parallel processing. 


In the gridding algorithm, visibilities are placed in parallel on a regular 2D grid of size $N_x \times N_x$ based on the $(u, v)$ coordinates of the associated baselines. Visibilities are not convolved with any kernel at this stage, a feature to be introduced in the near future to alleviate the effects of source aliasing. Natural weighting, where contributions from different baselines carry the same importance, is the default weighting scheme adopted to maximise S/N.

Once ready, gridded visibilities are passed to the ROCm or CUDA GPU-accelerated FFT library to compute the final images. The FFT implementation places the phase centre in the top-right corner of the image and the resulting images are shifted for a natural visual result. This operation is not necessary to find FRBs, and can be disabled, but its computational cost is negligible with respect to other steps.
 

\subsection{Running the pipeline}

The BLINK project makes available a command-line interface executable to run the pipeline described in this paper, aptly named \texttt{blink\_pipeline}. It takes as input a list of \texttt{.dat} files representing one or more seconds of a single MWA VCS observation, and several options to set parameters such as time resolution, fine channel bandwidth, image size, and calibration solutions. The returned output currently consists of FITS files containing an image for each frequency channel and integration interval. Ultimately, the resulting image cubes will be searched for FRBs.

A single executable implementing the entire pipeline also provides a an intuitive interface for researchers to use. An example of submitting a BLINK imaging job to Setonix is shown in Listing \ref{lst:submission_script}.  In contrast, established MWA imaging pipelines are made up of several BASH or Nextflow scripts chaining together a number of executable programs, each with its own input and output formats, and a wide range of command line options. The ease of use of the BLINK pipeline reduces the human time required to setup the computation, minimises configuration errors, script development time,  and improves accessibility to beginners in the field.

\subsection{Multi-vendor support for AMD and NVIDIA GPUs}
	While NVIDIA has been dominating the GPU market for decades, the adoption of AMD GPUs by several supercomputing centres has drastically increased their relevance in scientific computing. As of November 2024, five of the top ten most powerful supercomputers feature AMD GPUs \autocite{top500_2024}. The HIP programming language is the AMD equivalent of the CUDA C++ language extension and it is made available as part of the AMD ROCm  ecosystem.
	
	The BLINK software supports both AMD and NVIDIA ecosystems with a single codebase. The program interacts with the GPU runtime API through a set of custom-defined macros that abstract away vendor-dependent signatures in CUDA or HIP API calls. This is possible because HIP was designed to closely mimic the CUDA interface. For instance, to allocate memory on GPU we use the macro \texttt{gpuMalloc($\cdots$)}, that expands to \texttt{hipMalloc($\cdots$)} or \texttt{cudaMalloc($\cdots$)} whether the software is compiled for an AMD or NVIDIA system, respectively. In principle, programs written with HIP can also be compiled for an NVIDIA GPU, provided both HIP and CUDA are installed on the system. However, by using macros the BLINK pipeline can be compiled with the NVIDIA compiler without installing HIP.
	
	GPU kernel language is almost the same across the two platforms. One significant difference is that HIP provides native support within the kernel code for a portion of the C++ standard library. For instance, in a HIP kernel one can utilise the \texttt{std::complex} type and its operators. The same is not possible in CUDA, where the \texttt{cuComplex} data type must be used instead. We resorted to writing our own custom \texttt{Complex} class with the most commonly used operators, supporting execution on both CPU and GPU.
	
	Finally, the HIP compiler is based on LLVM Clang, whereas CUDA provides its own proprietary compiler driver that internally resorts to GNU \texttt{g++} to compile C++ code. These compilers provide different levels of support for the C++ language, and display different behaviours when programming errors or bad practices are detected. During the development, the code must be regularly compiled with both CUDA and HIP to make sure compatibility is maintained.

\section{Testing}\label{sec:testing}

Firstly, the BLINK imaging pipeline must produce reasonable dirty images of the radio sky to prove the overall correctness of the software. Secondly, it has to generate millisecond time resolution images with enough sensitivity to detect bright transients. The first objective is ensured by implementing unit tests for all major functions and classes. Low-level components are tested with synthetic data while the output of major computational routines, such as correlation and gridding, are compared to the output produced by the corresponding procedures in the reference software, the SMART imaging pipeline. Finally, the ability to detect bright pulses is tested by imaging in 20\,ms time resolution few seconds of a MWA observation containing a known pulsar.

\subsection{Testing environment}\label{sec:test_env}

Experiments have been run on two Pawsey supercomputing systems, Garrawarla and Setonix, and an internal CSIRO cluster named Virga. The first one was a cluster with 78 nodes, each providing one NVIDIA Tesla V100 32GB GPU, two Intel Xeon Gold 6230 20-core CPUs, and 384\,GB of memory, for a total compute capacity of 0.7\,PFLOPS. Garrawarla was a system dedicated to radio astronomers and was decommissioned at the beginning of 2025. Setonix is the new Pawsey supercomputer dedicated to general research, and replaced Garrawarla as supporting system for radio astronomy research. 

Setonix GPU partition contains 192 nodes, each with four AMD MI250X GPUs, a AMD Trento 64-core CPU, and 256\,GB of host memory. The Trento CPU delivers  2.51\,TFLOPS  consuming up to 280\,W of power.  Each MI250X GPU is made of two Graphics Compute Dies (GCDs) seen by the runtime as two separate GPUs. A MI250X has a total of 47.9\,TFLOPS peak compute capacity and Thermal Design Power (TDP) of 500\,W. 

At the time experiments were executed, both Garrawarla and Setonix mounted the same \texttt{/scratch} Lustre filesystem. Furthermore, GPU nodes of both systems have a local NVMe disk processes can use for fast I/O. In these experiments, input data is always read from \texttt{/scratch}. Output images are written to either the Lustre filesystem or the NVMe disk depending on the volume of data produced.

For both test cases the SMART pipeline was run on Garrawarla because the offline correlator requires NVIDIA CUDA, whereas the BLINK pipeline was executed on Setonix GPU nodes. While the latter can be compiled for an NVIDIA system, the current prototype targets MI250X specifications and demands GPUs with more than 32\,GB of memory to store high time resolution images, as shown in Table \ref{tab:data_volumes}. In the future, Managed Memory, currently adopted to ease the ingest of voltage data, might be implemented throughout the pipeline to ease GPU memory restrictions. Then, the two pipelines run on different hardware, and in a different modality. The SMART pipeline makes heavy use of CPU cores whereas BLINK only uses CPU to read in data and to orchestrate GPU computation. 

To abstract away these differences in parts of the discussion, we define the concept of execution unit as the minimal hardware requirements to execute an imaging pipeline. On Garrawarla, it is an entire GPU node. On Setonix, it corresponds to a single MI250X GCD and 8 CPU cores, or 1/8\textsuperscript{th} of a GPU node. Each instance of a pipeline is allocated an execution unit to process one or more seconds of observation sequentially. An observation can be partitioned in batches of seconds which are processed in parallel over multiple execution units.

This work focuses on the novel software architecture adopted by BLINK. However, to support the claim that the software supports both NVIDIA and AMD GPUs, in the second test case BLINK was also run on the Virga cluster after Garrawarla had been decommissioned. Virga is a GPU cluster featuring NVIDIA H100 GPUs.

\subsection{Long exposure image}

%

\begin{table}
	\begin{tabular}{|l|l|}
		\toprule
		\textbf{Property} & \textbf{Value} \\
		\bottomrule
		Observation ID (OBSID) & 1276619416 \\\hline
		MWA Configuration & Phase II Extended \\\hline
		Target & PSR J1834-0426 \\\hline
		UTC Start & 2020-06-19 16:29:58 \\\hline
		Duration (s) & 5400 \\\hline 
		Frequency range (MHz) & [169.6,   200.32] \\\hline
	\end{tabular}
	\caption{\textbf{Observation 1}. Information summary of the MWA observation imaged in a long exposure setting with the BLINK and SMART pipelines.}
	\label{tab:obs1_info}
\end{table}

The MWA observation OBSID 1276619416, details summarised in Table \ref{tab:obs1_info}, is adopted as a test case to demonstrate the ability of the BLINK imaging pipeline to produce correct dirty images starting from voltage data. The Extended configuration has longer baselines resulting in high angular resolution. The phase centre is situated in proximity of the galactic plane and generating a long exposure image gives a FoV with plenty of radio sources. The combination of both allows to visually diagnose coding errors and discern subtle differences in images produced by the BLINK and the SMART pipelines, mostly due to enabled features, completeness of the implementation, and design decisions. 

For this test the following is performed with both pipelines. Voltages corresponding to the first 4660 seconds of the observation are correlated at 1\,s time resolution and the fine channel bandwidth is 40\,kHz. For each second of data and for each fine channel visibilities are gridded into an array of 8192x8192 pixels and then Fourier transformed. The resulting 1\,s images are then averaged in time over the entire processed time frame and frequency over the 30.72\,MHz bandwidth, resulting in one long exposure image. For both pipelines, I/O operations are directed to the \texttt{/scratch} filesystem given the large volume of data written during the processing of 1.5h of observation.

 The relative difference between visibilities produced by SMART and BLINK is in the order of 0.001\%. It is mainly during the process of gridding that differences in the BLINK imager and WSClean implementations, such as the absence of a gridding kernel and less flexibility in setting the pixel scale in the first one,  lead to visually identifiable discrepancies between the resulting images.

\begin{table}[h!]
	\begin{tabular}{|l|c|c|}
		\toprule
		\textbf{Property} & \textbf{BLINK image} & \textbf{SMART Image} \\\bottomrule
		
		Top region noise (Jy)              & 0.32 & 0.28 \\\hline
		Centre region peak flux (Jy) & 1.79 & 1.96 \\\hline
		Bottom region peak flux (Jy) & 5.1 & 4.35 \\\hline
		
	\end{tabular}
	
	\caption{\textbf{Key image statistics.} This table reports key statistics of the three regions highlighted in the long exposure images produced by the BLINK and SMART pipelines. The top region does not contain any source and it is ideal to sample the noise level. The two other regions contain extended objects, providing opportunity to compare peak fluxes.}
	\label{tab:test1_stats}
\end{table}

\begin{figure*}[ht]
	\centering
    \setkeys{Gin}{width=0.4\linewidth} 
	\subfloat[BLINK image \label{fig:blink_1hour}]{\includegraphics{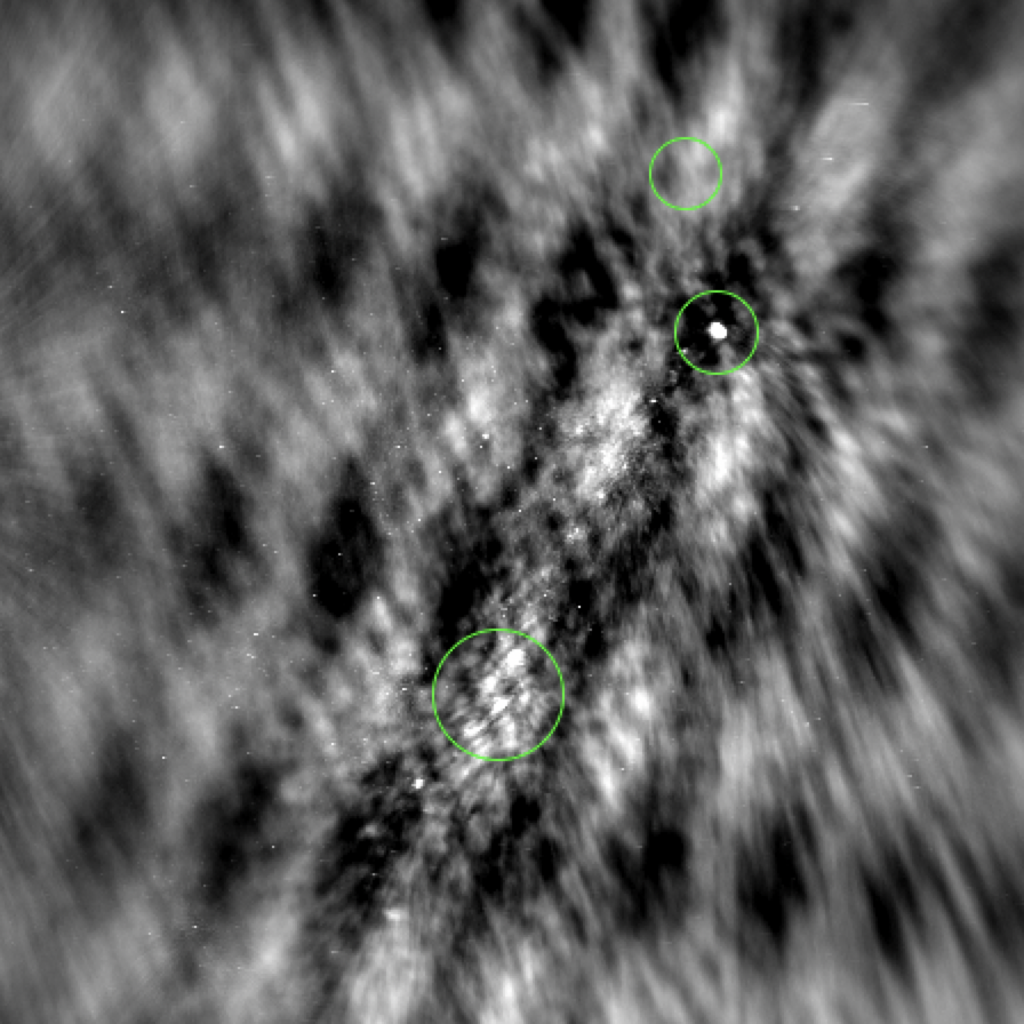}} \hspace{2em}
	\subfloat[SMART image \label{fig:wsclean_1hour}]{\includegraphics{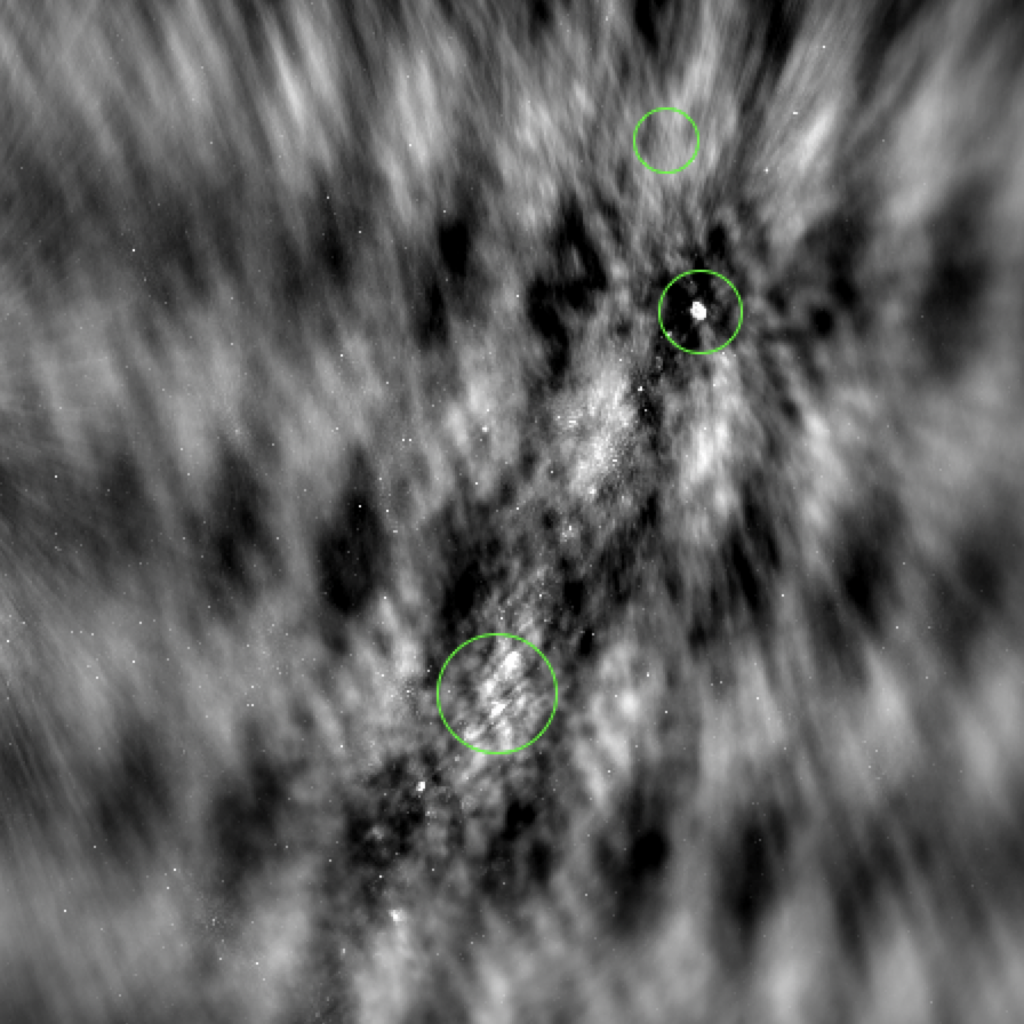}}\hfil
	
	\caption{\textbf{Long exposure images}. The BLINK and SMART pipelines were run on the MWA Phase II Extended 1276619416 observation in multi-frequency synthesis mode, averaging images across all time bins, which resulted in the long exposure images shown in these panels. Three regions are highlighted by green circles in both images to compute key statistics for the purpose of comparison. These are reported in Table \ref{tab:test1_stats}.}
	\label{fig:long_exposure_image}

\end{figure*}

Images produced by the BLINK and WSClean imagers are shown in panels (a) and (b), respectively, of Figure \ref{fig:long_exposure_image} as rendered by the ds9 software with the z-scale option enabled. At first glance they look almost identical. At the centre of the image is the galactic plane showing multiple bright sources. We highlight three regions of the sky with green circles. The first one, at the centre top, contains no source and we measure its noise. The centre region highlights an extended source. The third one, at the bottom, encloses a supernova remnant. Striping patterns are most likely due to undeconvolved sidelobes of the complicated galactic plane in the image.

A difference image could not be created due to differing pixel sizes in the two images. Instead, corresponding regions are compared in terms of noise or peak flux to quantify differences, and results are presented in Table \ref{tab:test1_stats}. Respective peak flux density and noise measures present a relative difference between $9\%$ and  $14\%$ . There are a few factors that contribute to the difference. For instance, the BLINK imager currently does not compute Stokes I and only employs the XX polarisation\footnote{WSClean is executed with IDG enabled which produces Stokes I images but not XX polarisation ones for comparison.}, nor it applies a gridding kernel to alleviate aliasing effects.

 
\subsection{High time resolution images}

\begin{table}
	\begin{tabular}{|l|l|}
		\toprule
		\textbf{Property} & \textbf{Value} \\
		\bottomrule
		Observation ID (OBSID) & 1293315072 \\\hline
		MWA Configuration & Phase II Extended \\\hline
		Target & PSR B0950+08  \\\hline
		UTC Start & 2020-12-29 22:10:54\\\hline
		Duration (s) & 340 \\\hline 
		Frequency range (MHz) & [138.8, 169.6] \\\hline
	\end{tabular}
	\caption{\textbf{Observation 2.} Details of the MWA observation that has been imaged at 20\,ms time resolution with the BLINK imager.}
	\label{tab:obs2_info}
\end{table}

The second test consists in the generation of high time resolution images of MWA OBSID 1293315072, details in Table \ref{tab:obs2_info}, to detect the pulses of PSR B0950+08. A single radio emission from a pulsar is the closest resembling transient event to an FRB that has been recorded with the MWA, and its detection would provide great confidence that the presented pipeline is sensitive to millisecond radio transients. 

The BLINK pipeline has been run on the first seven seconds worth of VCS data by correlating voltages at 20\,ms time and 40\,kHz frequency resolutions. This configuration results in 50 time bins and 768 fine channels. Visibilities from each frequency channel and time bin are gridded in parallel on separate grids. The process produces 38400 images of chosen resolution 1200x1200 pixels per second of observation, then written to a node-local NVMe volume for visual inspection. Images belonging to the same time bin are averaged together, effectively covering the whole 30.75\,MHz frequency bandwidth of the MWA. The dispersive delay affecting the signal is almost negligible because the DM is only 2.97, and the de-dispersion is not applied to simplify this demonstrative analysis.

\begin{figure*}[ht]
	\begin{center}
		\includegraphics[width=0.90\textwidth,angle=0]{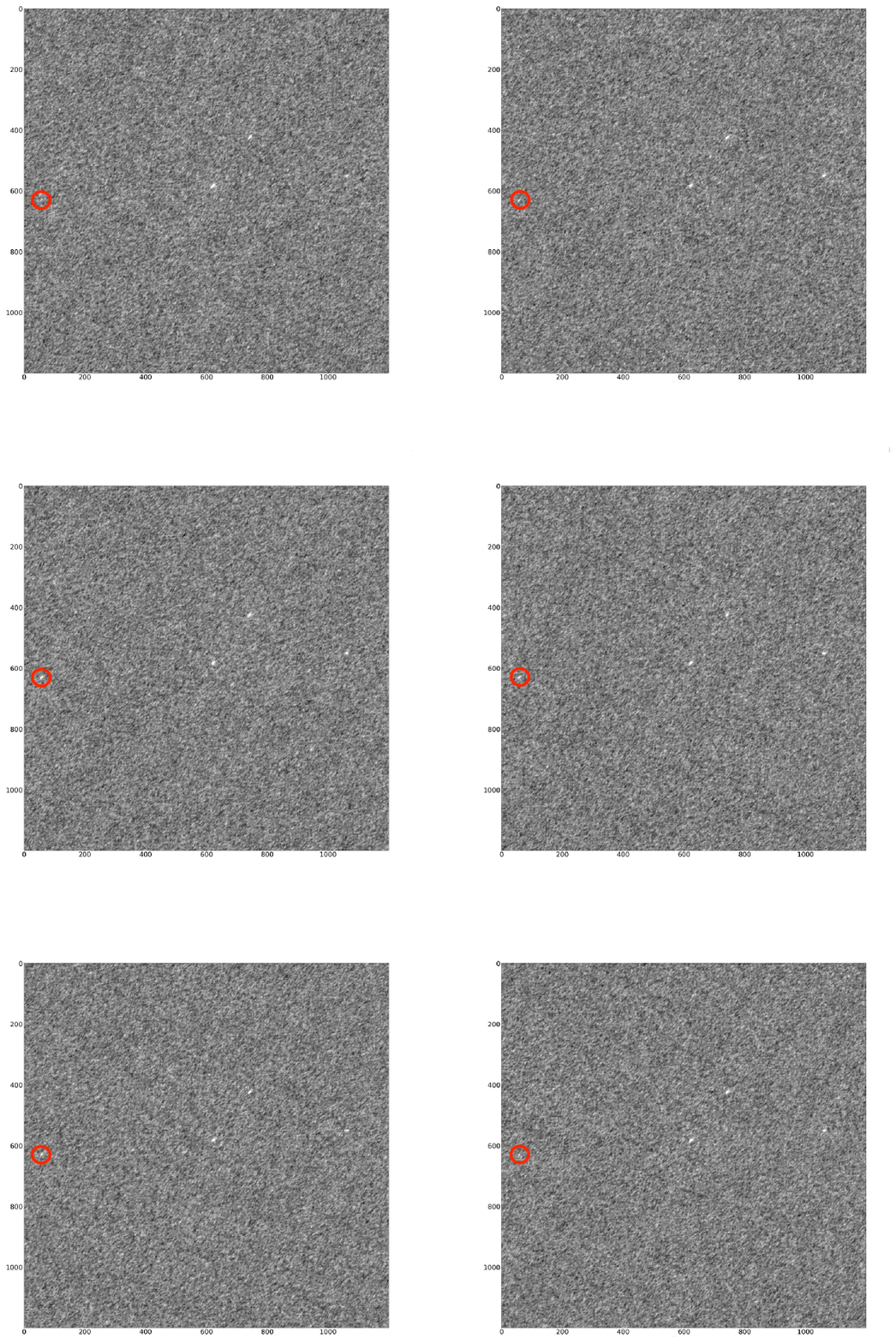} 
		
		\caption{\textbf{High time resolution images.} The figure shows a sample of the 20\,ms images generated with the BLINK pipeline as part of the second test case. These contain bright pulses of pulsar B0950+08, whose position is marked with a red circle. The centre row shows the images with bright single pulses at UNIX times 1609279857.740 (left column) and 1609279856.480 (right column). The 20\,ms images before and after the pulses are respectively in the rows above and below the centre one. When the peak of the pulse is closer to the edges of time bins its residual signal appears in the before and after images. The average noise in the images is equal to 0.5\,Jy. The two brightest pulses have a flux density of 9 and 16.5\,Jy.}
		\label{fig:B0950_20ms_panel}
	\end{center}
\end{figure*}

The radio pulse of PSR B0950+08 is clearly visible in the image sequence, a sample of which is shown in Figure \ref{fig:B0950_20ms_panel}. Visibilities are not phase-centred at the location of the pulsar, then seen in the middle-left part of the FoV and highlighted by a red circle in the images. Each column depicts an independent pulse detection. Images in the centre row capture the peak of pulses, whereas the first and third show the sky intensity moments before and after those. The noise level sits at 0.5\,Jy, which is in good agreement with the expected value based on electro-magnetic simulations. The two brightest pulses have a flux density of 9 and 16.5\,Jy.

We attempted to run the SMART pipeline in an equivalent configuration but WSClean could only generate 2730 of the 38400 images within the 24 hours wall time imposed on Pawsey supercomputing systems. This implies an execution time that is too long for WSClean to efficiently image multiple frequency channels and time intervals over entire observations.
%
%
%

\section{Benchmarking}\label{sec:benchmarking}


The computational performance of the BLINK pipeline against the SMART one is evaluated in terms of execution time needed to process one second of full-bandwidth observation given a single execution unit as defined in Section \ref{sec:test_env}. We report the timings for both test cases although it is the second test that is representative of the compute requirements of the target science case.  Timings of the imaging stages are recorded for each second of observation processed. However, for the SMART pipeline it was not possible to distinguish between time spent in I/O and computation at each stage due to the fact these operations interleave. In the BLINK pipeline input ingestion occurs at the beginning before any computation starts. We also report the time it takes to write out the final images, for completeness. That last operation will not be required when the pipeline will be extended to support FRB searches.

Test case one processed 4660 seconds of observation partitioned in batches of 100, each batch submitted as an independent computation to the supercomputer. Timings recorded for each second are averaged and reported in Table \ref{tab:test1_profile}. On average, the BLINK pipeline is 3.8 times faster than the SMART one, with mean total execution times, respectively, of $48.75$ and $184$ seconds. Close to $95\%$ of the BLINK execution time is spent in reading input files and writing out images, operations whose distribution of timings is characterised by a significant standard deviation. While we could not directly quantify it, I/O activities impact the SMART pipeline as demonstrated by the mean execution time and the standard deviation of the corrections step. This is implemented by the COTTER program that writes its output as a CASA MeasurementSet. It has been experimentally observed that writing data to such format incurs in degraded performance on Pawsey's \texttt{/scratch} Lustre filesystem, mainly due to the large number of requests to metadata servers. The contribution to the total execution time of the computational kernel is minimal in comparison. Hence, the GPU acceleration provides little improvement in this mode of operation.

The advantage of the BLINK pipeline over the classic MWA imaging pipeline becomes apparent when moving to high time and frequency resolution imaging. Table \ref{tab:test2_profile} reports timings of the two imaging pipelines when processing 1 second worth of data at 20\,ms time and 40\,kHz frequency resolutions. The execution time of the SMART pipeline is completely dominated by WSClean that does not parallelise the task of generating images for multiple frequency channels and time bins. For this reason it only generated $7\%$ of the expected total number of images within the 24 hours wall time limit on Garrawarla. Based on that information, it would take 14 days for WSClean to complete on a single Garrawarla node. One could think of running multiple WSClean instances on many nodes but the cost in terms of node-hours would not change.

The BLINK pipeline performs the same task in 300 seconds using a single MI250X GCD, achieving a 3687x speed up. Furthermore, $87\%$ of the total execution time, corresponding to 262\,s, is spent writing images to the NVMe disk, a step that will not be necessary during FRB searches. Hence, the current implementation of the BLINK imaging pipeline is 38 times slower than real-time for the proposed test case. The two most computationally expensive steps are correlation and 2D FFT, as expected, executing in 6.9\,s and 6.46\,s respectively. In particular, the 2D FFT routine processes 38,400 input grids using batched GPU computation, and its runtime is only twice as long as the version that handles a single grid in the long exposure test case.

The BLINK pipeline was also run on the Virga cluster using a single NVIDIA H100 GPU and results are reported in Table  \ref{tab:test2_profile}. The total runtime is almost three times higher than the one measured on Setonix due to slower I/O. However, GPU computations such as gridding and correlation are significantly faster, with execution times of 0.09\,s and 2.01\,s respectively.

Finally, we report a total energy consumption of 30597\,J, or an average power consumption of 98\,W by the MI250X GPU to run the high time resolution test case. The power usage is well below the TDP of the MI250X, signalling underutilisation. This is to be expected as only one of the two GCDs is being used, and most of the execution time of the test case is spent writing images to disk.

\begin{table*}
	\centering
\begin{tabular}{|l|c|c|c|c|c|c|c|}
	\toprule
	\textbf{Pipeline} & \textbf{Total time (s)} & \textbf{Data loading (s)} & \textbf{Correlation (s)} & \textbf{Corrections (s)} & \textbf{Gridding (s)} & \textbf{2D FFT (s)} & \textbf{Output writing (s)} \\
	\bottomrule
	BLINK & 48.75 (2.60) & 40.76 (6.85) & 2.67 (0.009) & 0.007 ($3.25 \cdot 10^{-5}$) & 0.009 ($2.43 \cdot 10^{-5}$) & 3.02 (0.42) &  4.51 (0.11) \\\hline
	SMART & 184 (331.8) & N.A.\footnote{I/O operations are present at every step of the SMART pipeline and their contribution to the total execution time cannot be clearly discerned from numerical computation. Hence, values in the other columns are inclusive of input reading and output writing activities.} & 25.9 (2.6) & 113.1 (331.8) &  \multicolumn{2}{c|}{44.9 (2.2) } & N.A. \\\hline
\end{tabular}

\caption{\textbf{Execution times for the long exposure images}. Execution times and their standard deviation (in brackets) of the various phases of the SMART and BLINK imaging pipelines needed to process one second of observation at 1\,s time resolution and 40\,kHz frequency resolution. Values are an average over 4660\,s worth of observation. For the SMART pipeline, timings for the Data loading and Output writing are not available because I/O interleaves with computation at every stage. The Correlation, Corrections, and Gridding and 2D FFT columns correspond to the execution of offline correlator, COTTER and WSClean, respectively.}
\label{tab:test1_profile}
\end{table*}

\begin{table*}
	\centering
	\begin{tabular}{|l|c|c|c|c|c|c|c|}
		\toprule
		\textbf{Pipeline} & \textbf{Total time (s)} & \textbf{Data loading (s)} & \textbf{Correlation (s)} & \textbf{Corrections (s)} & \textbf{Gridding (s)} & \textbf{2D FFT (s)} & \textbf{Output writing (s)} \\
		\bottomrule
		BLINK MI250X & 300 & 6.2 & 6.9  & 2.55 & 0.626 &  6.46 &  262  \\\hline
		BLINK H100 & 840 & 37 & 2.01  & 1.59 & 0.09 &  1.68 &  761  \\\hline		
		SMART & 121.68 $\cdot$ 10$^4$ (projected) \footnote{This is equal to 338 hours, or 14 days.} & N.A. & 144 & 107 &  \multicolumn{2}{c|}{121.68 $\cdot$ 10$^4$(projected) } & N.A. \\\hline
	\end{tabular}
	
	\caption{\textbf{Execution times for high resolution images.} The timing for each step of the SMART and BLINK imaging pipelines needed to process one second of observation at 20\,ms time and 40\,kHz frequency resolutions. In this case WSClean could only complete 7\%  of the total number of imaging steps within the 24h wall time limit of Garrawarla. Hence, we give a projected execution time based on that information. No standard deviation is reported given timings of kernel executions and I/O operations on NVMe are quite stable.}
	\label{tab:test2_profile}
\end{table*}

\begin{table*}
	\centering
	\begin{tabular}{|l|c|c|c|}
		\toprule
		\textbf{Test case} & \textbf{Voltages} & \textbf{Visibilities} & \textbf{Images} \\
		\bottomrule
		Long exposure image & 7.5\,GB & 194\,MB & 256\,MB \\\hline
		High resolution images & 7.5\,GB & 9.5\,GB & 206\,GB \\\hline
	\end{tabular}
	
	\caption{\textbf{Data volumes.} A summary of the amount of memory required to hold fundamental data products generated at each step of an imaging pipeline for the two test cases discussed in this paper. The table does not account for additional support data structures such as the array holding gridded visibilities.}
	\label{tab:data_volumes}
\end{table*}


\section{Discussion and conclusions}\label{sec:discussion}

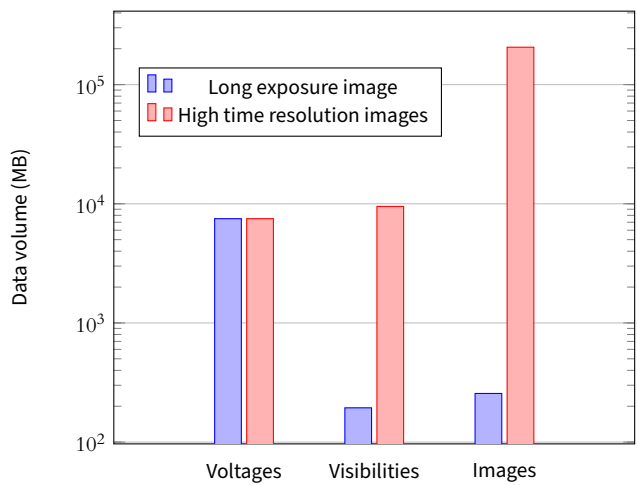
\begin{figure}[ht!]
	\begin{tikzpicture}
		\begin{axis}[
			major x tick style = transparent,
			ybar,
			ymode=log,
			enlarge x limits=0.5,
			ymajorgrids = true,
			xtick = data,
			ylabel = {Data volume (MB)},
			legend style={at={(0.63,0.87)}},
			symbolic x coords={Voltages, Visibilities, Images}]
			\addplot
			coordinates {
				(Voltages,7500)(Visibilities,194)(Images,256)
			};
			\addplot
			coordinates {
				(Voltages,7500)(Visibilities,9500)(Images,206000)
			};
			\legend{Long exposure image, High time resolution images}
		\end{axis}
	\end{tikzpicture}
	\caption{\textbf{Data volumes generated by an imaging pipeline per second of observation.} In blue is the case of multi-frequency synthesis of a single image using full bandwidth data and a long (1\,s) integration time. The red bars shows data volumes in the case of high time (50 $\times$ 20\, ms time bins) and frequency (768 $\times$ 40\,kHz channels) resolution images. No significant averaging in time and frequency occurs in the latter case, resulting in a large number of images being generated.}
	\label{fig:data_rates}
\end{figure}

Imaging could become the predominant technique for wide-field interferometers like the MWA to detect and localise transients such as FRBs due to its theoretical computational advantage over beamforming. However, its implementation has to efficiently manage the processing and storage of the large number of images resulting from the millisecond time and kHz frequency resolution requirements. Figure \ref{fig:data_rates} illustrates the data volume involved at each step of the imaging pipeline for the long exposure and high time and frequency resolution cases. Actual numbers are reported in Table \ref{tab:data_volumes}. In high time and frequency resolution imaging data volumes become larger at each further stage, as opposed to the traditional long exposure imaging case.

Execution times of established applications that make up the SMART pipeline are heavily affected by I/O operations and the performance variability of the Lustre parallel filesystem, as shown in Table \ref{tab:test1_profile} and Table \ref{tab:test2_profile}. BLINK addresses the performance and software design limitations imposed by I/O operations performed at every stage of the traditional imaging pipeline. It does so by adopting an in-memory processing philosophy where data resides in the GPU memory at all times. Input reading occurs at the beginning of the pipeline and it is throttled depending on the availability of GPU memory. This strategy works well because each second of observation can be processed independently. In contrast, when data is staged in host memory or even disk at each step of the pipeline, time consuming memory transfers between the CPU and GPU memory spaces are necessary to provide the input to GPU kernels and to retrieve the produced output.

Processing speed becomes also a concern with petabytes worth of observations to analyse. Every computational stage of the imaging process is amenable to a parallel implementation. The BLINK pipeline adopts the GPU not as an accelerator but as main compute hardware device. Input voltages are transferred to GPU memory and a sequence of compute kernels are scheduled for execution, in many cases without the need to synchronize with the CPU. Each kernel operates in parallel on multiple data dimensions: array baselines, integration intervals, and frequency channels. This design enables massive parallelism by running the GPU at maximum capacity by feeding it as much data as it can fit into memory. The BLINK pipeline is not only over three orders of magnitude faster than the SMART one, but also more energy efficient. A MI250X performs more operations than a Trento CPU consuming the same amount of power: 95 GFLOPS per Watt compared to 9\,GFLOPS per Watt, respectively.

In conclusion, the BLINK imaging pipeline presented in this work provides a fast, scalable , and energy efficient way to image MWA observations in millisecond-time and kHz frequency resolutions using the Setonix supercomputer. The produced dirty images are a suitable input to a search pipeline with the objective of detecting and localising fast transients like FRBs.




\begin{lstlisting}[frame=single, float=*, language=bash,basicstyle=\ttfamily,
	caption={\textbf{Example submission script.} Executing the BLINK pipeline on the Setonix supercomputer requires minimal BASH scripting. First, Slurm directives inform the scheduler of the computational resources required to run the program. In this case, a single GPU on a node in the \texttt{gpu} partition. BASH variables are then defined to point to input files like observation metadata, calibration solutions, and voltage files. BASH wildcards are used to select the interval of seconds and frequency channels to process. In this case, all the data associated with OBSID 1293315072. Finally, the BLINK pipeline is executed with a single command line. Important options are:  \texttt{-t}, the integration time; \texttt{-c}, the number of contiguous fine channels to average; \texttt{-n}, the side size in pixels of the output images.}\
	\
	]
	#!/bin/bash -e 
	#SBATCH --gres=gpu:1
	#SBATCH --partition=gpu
	#SBATCH --time=24:00:00
	#SBATCH --export=NONE
	#SBATCH --output=logs/slurm-%A.out 
	
	INPUT_DIR=/scratch/mwavcs/cdipietrantonio/1293315072/combined
	INPUT_FILES="${INPUT_DIR}/1293315072_*_ch*.dat"
	METAFITS=1293315072.metafits
	SOLUTIONS_FILE=1141222488.bin 
	
	OUTPUT_DIR=$MYSCRATCH/1293315072_int20ms_1200px

	
	blink_pipeline -c 4 -t 20ms -o ${OUTPUT_DIR} \
	    -n 1200 -M ${METAFITS} -w N -s ${SOLUTIONS_FILE} ${INPUT_FILES} 
\end{lstlisting}\label{lst:submission_script}

\section{Future work}\label{sec:future_work}

The next objective for the authors is to extend the pipeline with additional modules implementing FRB search techniques. However, much can be done to the current BLINK imaging pipeline to improve the quality of the output. The implementation of a gridding kernel to minimise aliasing is in progress. Another task is using both XX and YY polarisations when gridding visibilities, instead of only the first one. The performance and stability of the implementation can be enhanced too by fine tuning the GPU kernels, implementing memory allocation reuse, experimenting with reduced precision, and more. In general, we welcome any code contribution to extend the capabilities of the BLINK imaging pipeline to support other telescopes and science cases other than FRBs.

\begin{acknowledgement}
This scientific work uses data obtained from Inyarrimanha Ilgari Bundara / the Murchison Radio-astronomy Observatory. We acknowledge the Wajarri Yamaji People as the Traditional Owners and native title holders of the Observatory site. Establishment of CSIRO's Murchison Radio-astronomy Observatory is an initiative of the Australian Government, with support from the Government of Western Australia and the Science and Industry Endowment Fund. Support for the operation of the MWA is provided by the Australian Government (NCRIS), under a contract to Curtin University administered by Astronomy Australia Limited. This work was supported by resources provided by the Pawsey Supercomputing Research Centre’s Setonix Supercomputer (\url{https://doi.org/10.48569/18sb-8s43}), with funding from the Australian Government and the Government of Western Australia. Authors also acknowledge the Pawsey Centre for Extreme Scale Readiness (PaCER) for funding and support.
\end{acknowledgement}

\appendix

\printendnotes

\printbibliography

\end{document}